\documentclass[showpacs,amsmath,amssymb]{revtex4}
\usepackage{graphicx}

\begin{document}

\title{Shape resonances in the superconducting order parameter of ultrathin nanowires}
\author{A. A. Shanenko,}
\affiliation{Bogoliubov Laboratory of Theoretical Physics, Joint Institute for Nuclear
Research, 141980 Dubna, Russia}
\author{M. D. Croitoru}
\affiliation{EMAT, University of Antwerp, Groenenborgerlaan 171, B-2020 Antwerp, Belgium}

\date{\today}

\begin{abstract}
We study the shape resonance effect associated with the confined transverse superconducting modes
of a cylindrical nanowire in the clean limit. Results of numerical investigations of the Bogoliubov-de
Gennes equations show significant deviations of the energy gap parameter from its bulk value with
a profound effect on the transition temperature. The most striking is that the size of the resonances
is found to be by about order of magnitude larger than in ultrathin metallic films with the same width.
\end{abstract}


\maketitle

Modern rapid miniaturization of electronic circuits requires good understanding of basic mechanisms
responsible for the electronic properties of nanoscale structures. The most important point about
these structures is that the quantum-confinement effects play the corner-stone role in this case.
One can even say in general that recent success in nanofabrication technique has resulted in great
interest in various artificial physical systems with unusual phenomena driven by the quantum
confinement (quantum dots, nanoscale semiconductors, nanosuperconductors, etc.). The quantum-confined
superconductivity is here of special interest due to the macroscopic quantum character: any effect on
electron wave functions manifests itself directly in the superconducting order parameter.

An obvious consequence of the confinement in a nanoscale superconducting structure is nonuniform
spatial distribution of the superconducting condensate because, as it is known since the classical
papers by Gor'kov~\cite{gor} and Bogoliubov~\cite{bog1}, the superconducting order parameter
can be interpreted as the wave function of the center-of-mass motion of a Cooper pair. It is also
known that the Cooper-pair wave function involves important in-medium terms~\cite{cooper}. In the
presence of the electron confinement these terms can result in shape resonances in the energy gap
parameter, another confinement effect first found and investigated in the paper by Blatt and
Thompson~\cite{blatt} for ultrathin metallic films. A shape resonance in the dependence of the
energy-gap parameter on the specimen dimensions can occur any time when an electron subband
appearing due to the size quantization passes through the Fermi surface~\cite{blatt}. Strong
indications for such behaviour are found not only in ultrathin films but also in nanoparticles
(see, for example, Refs.~\cite{farine} and \cite{iv}) and superfluid nuclei~\cite{satula,hilaire}
(as it had been predicted by Blatt and Thompson). Recently a new technique of electrodeposition into
extended nanopores has been developed~\cite{tian}, which makes it possible to produce single-crystal
nanowires of high quality. Thus, the shape resonances in the nanowire superconducting order parameter
can be investigated in the clean limit, with a direct link to the microscopic (BSC) theory. In
particular, it is of importance to explore the situation where the QPS (quantum phase slips)
regime~\cite{tian,gior,sharifi,ling}) is expected to generate a new low-temperature metallic phase
with proliferating quantum phase slips of the superconducting order parameter (for radii less than
$5\;nm$)~\cite{zaik}. The point is that when calculating the QPS energy barrier~\cite{tian, zaik}
in ultrathin nanowires, one assumes the superconducting order parameter being uniform in the transverse
direction but with the phase slips in the longitudinal one~\cite{zaik}. The absolute value of
the order parameter is set to be equal to the bulk one. However, at a resonant point the
superconducting condensate shows significant spatial variations in the transverse direction, and
its mean absolute value can be much larger than that of the bulk material. Note that the quantum
confinement can influence the superconducting state by two channels. The first is due to change
in electron wave functions (this is a main reason for the shape resonance effect). Whereas the
second is connected with the confined phonons. The both channels were examined for ultrathin
metallic films~\cite{hwang}, which makes it possible to expect that confinement modifications
of the phonon modes can produce quantitative corrections if the nanowire width is less than or
about $2\;nm$. In the present work the first channel is under consideration, while the phonons
are taken to be the same as in the bulk material. Thus, below we investigate the shape resonance
effect associated with the confined transverse superconducting modes of an ultrathin nanowire
in the clean limit.

\begin{figure}[t]
\centerline{\includegraphics[width=14.6cm,clip=true]{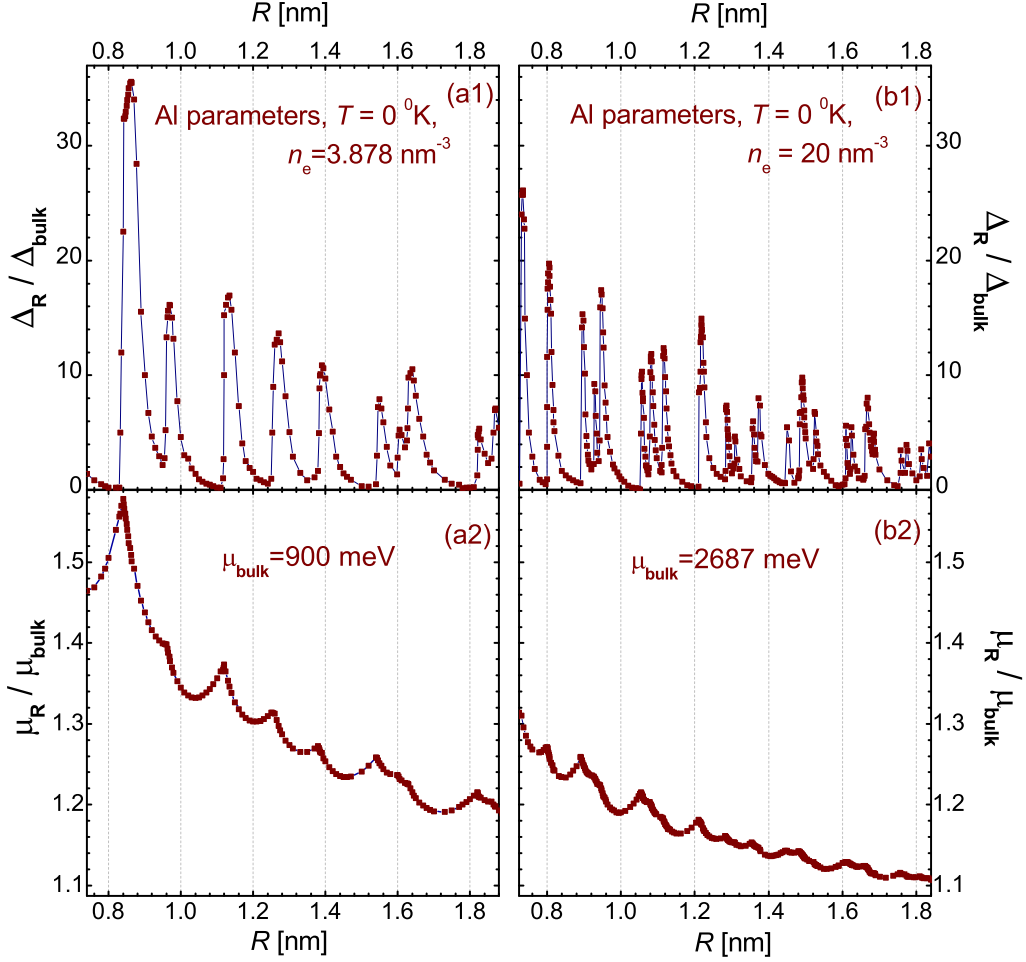}}
\vspace{-0.8cm}
\caption{The relative gap of the quasiparticle spectrum $\Delta_R/\Delta_{bulk}$ (panels (a1)
and (b1)) and relative chemical potential $\mu_R/\mu_{bulk}$ (panels (a2) and (b2)) versus the
nanowire radius $R$: the panels (a1) and (a2) represent the data calculated for $n_e=3.878\;
nm^{-3}$ (the bulk chemical potential $\mu_{bulk}=900\;meV$); the panel (b1) and (b2) are the
data for $n_e = 20\;nm^{-3}$ ($\mu_{bulk} = 2687\;meV$). Squares are the results of numerical
investigations of the BdG equations, the solid line is the spline interpolation.}
\label{fig1}
\end{figure}

To explore the superconducting order parameter varying with position, one should use the
Bogoliubov-de Gennes (BdG) equations~\cite{degen}. We are interested in numerical solutions of BdG
equations taken in the absence of magnetic field for a superconducting cylinder with the radius $R$
and length $L$. In all the calculations $L$ remains the same and equal to $2000\;nm$ while $R$ is
varied from $1\; nm$ to $10\;nm$. In the absence of magnetic field the superconducting order
parameter $\Delta({\bf r})$ can be chosen as a real quantity (phase effects are beyond the scope
of our consideration) and the BdG equations have the form
\begin{eqnarray}
E_n u_n({\bf r})= \Bigl(-\frac{\hbar^2}{2m^*} \nabla^2 -\mu\Bigr)u_n({\bf r}) + \Delta({\bf r})
                                                                                     v_n({\bf r}),
\label{BdG1}\\ [1mm]
E_n v_n({\bf r})= \Delta({\bf r}) u_n({\bf r})- \Bigl(-\frac{\hbar^2}{2m^*} \nabla^2 -\mu\Bigr)
                                                                                     v_n({\bf r}),
\label{BdG2}
\end{eqnarray}
where $E_n$ stands for the quasiparticle spectrum, $\mu$ is the chemical potential and $m^*$ denotes
the electron band mass. The single electron wave functions $u_n$ and $v_n$ make a contribution to
the order parameter via the self-consistency relation
\begin{equation}
\Delta({\bf r})= g \sum\limits_n u_n({\bf r})v_n^*({\bf r}) \bigl(1 - 2f(E_n)\bigr),
\label{slf_rel}
\end{equation}
where $g$ is the coupling constant and $f(x)$ is the Fermi function $f(x)=1/(exp(\beta x) + 1)$,
$\beta=1/(k_B T)$ with $T$ the temperature and $k_B$ the Boltzmann constant. Summation in
Eq.~(\ref{slf_rel}) is taken over the eigenstates with the kinetic energy (including the chemical
potential) within the window $[-\hbar\omega_D, \hbar \omega_D]$, and $\omega_D$ is the Debye
frequency. The chemical potential is fixed by
\begin{equation}
n_e = 2 \sum\limits_n \left[|u_n({\bf r})|^2 f(E_n)+|v_n({\bf r})|^2 (1-f(E_n))\right]
\label{dens}
\end{equation}
with $n_e$ the total electron density. Introducing the cylindrical coordinates $\rho, \varphi, z$,
we can write $\Delta({\bf r})= \Delta(\rho)$ as the periodical boundary conditions are implied in
the longitudinal (z) direction. In this case we get
\begin{equation}
u_{jmk}({\bf r}) =  \widetilde{u}_{jmk}(\rho)\, \frac{e^{i m \varphi}}{\sqrt{2 \pi}}\,\frac{e^{ikz}}
{\sqrt{L}}, \;\;
v_{jmk}({\bf r}) =  \widetilde{v}_{jmk}(\rho)\, \frac{e^{i m \varphi}}{\sqrt{2 \pi}}\,\frac{e^{ikz}}
{\sqrt{L}}, \label{factor}
\end{equation}
where $n=(j,m,k)$ with $j$ the quantum number associated with $\rho-$coordinate, $m$ the azimuthal
quantum number and $k$ the wave vector in z-direction. Substituting Eq.~(\ref{factor}) into
Eqs.~(\ref{BdG1}) and (\ref{BdG2}), we recast the BdG equations in terms of $\widetilde{u}$ and
$\widetilde{v}$:
\begin{eqnarray}
E_{jmk}\; \widetilde{u}_{jmk}(\rho)= \left[-\frac{\hbar^2}{2m^*}\Bigl( \frac{\partial^2}{\partial
\rho^2}+\frac{1}{\rho}\frac{\partial}{\partial \rho} -\frac{m^2}{\rho^2} - k^2\Bigr) -\mu\right]
\widetilde{u}_{jmk}(\rho) + \Delta(\rho)\widetilde{v}_{jmk}(\rho),
\label{BdG_simp1}\\ [1mm]
E_{jmk}\; \widetilde{v}_{jmk}(\rho)= \Delta(\rho) \widetilde{u}_{jmk}(\rho)-
\left[-\frac{\hbar^2}{2m^*}\Bigl( \frac{\partial^2}{\partial \rho^2}
+\frac{1}{\rho}\frac{\partial}{\partial \rho} -\frac{m^2}{\rho^2} - k^2\Bigr) -\mu\right]
                                                                    \widetilde{v}_{jmk}(\rho).
\label{BdG_simp2}
\end{eqnarray}
Due to the electron confinement in the transverse directions we should put
\begin{equation}
\widetilde{u}_{jmk}(R)=\widetilde{v}_{jmk}(R)=0.
\label{conf}
\end{equation}
Then, $\widetilde{u}-$ and $\widetilde{v}-$functions are expanded in terms of the Bessel functions,
and Eqs.~(\ref{BdG_simp1}) and (\ref{BdG_simp2}) can be converted into a matrix form convenient for
the numerical iteration procedure typical for a self-consistent mean-field treatment.

\begin{figure}[t]
\centerline{\includegraphics[width=12.6cm,clip=true]{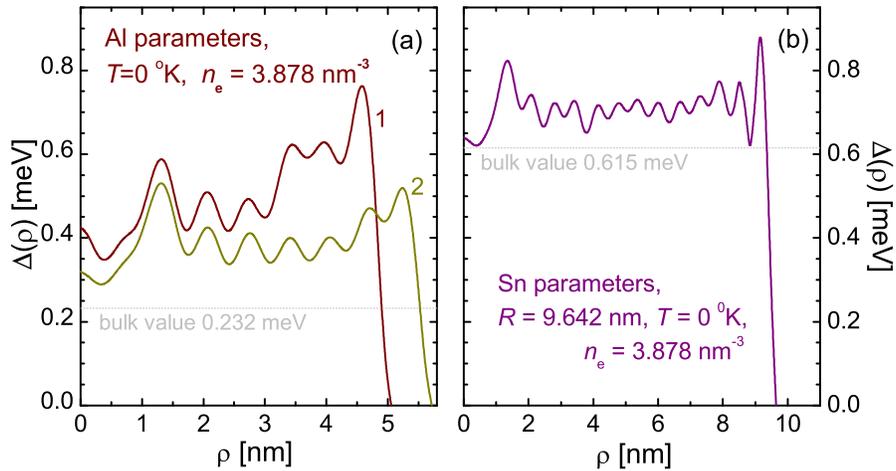}}
\vspace{-0.7cm}
\caption{The superconducting order parameter $\Delta(\rho)$ versus $\rho$ at $n_e=3.878\;nm^{-3}$ and
$T=0$: (a), the resonance points $R=5.06\;nm$~(1) and $R=5.75\;nm$~(2) for the $Al$ parametric set;
(b), the resonance point $R=9.642\;nm$ for the $Sn$ parameters.}
\label{fig2}
\end{figure}
Numerical investigation of Eqs.~(\ref{BdG_simp1}) and (\ref{BdG_simp2}) reveals the resonance
structure of the energy gap parameter of a cylindrical nanowire in agreement with the results
for ultrathin films and nanoparticles. But what is surprising is that the resonance size is now
by order of magnitude larger than in ultrathin films of the same width (see the results for
ultrathin metallic films with the width $10-20 \;\AA$~\cite{blatt}). In Fig.1 the $R-$dependent gap
of the quasiparticle spectrum $\Delta_R$ together with the chemical potential $\mu_R$ are given
for $ 0.8\;nm \leq R \leq 1.8\; nm$ at the zero temperature. The $Al-$parameter set was used in
the calculations: $g N(0)=0.18$~($N(0)$ is the bulk energy density of states for one spin projection
at the Fermi surface) and $\hbar \omega_D/k_B = 375\,^o\!K$~\cite{fetter}. The electron band mass
was set to the free electron mass. To have a feeling about influence of the total electron density
on the shape resonances, we performed calculations for the densities $n_e=3.878\;nm^{-3}$ (with
the bulk chemical potential $\mu_{bulk}=900\;meV$) and $n_e= 20\;nm^{-3}$ (with $\mu_{bulk}=2867\;meV)$.
Any time when the number of electron subbands (associated with the quantum numbers $j$ and
$m$) below the Fermi level gets less by $1$ or $2$ with decrease of the wire radius $R$
($1$ for the situation of $j$ dropping, $2$ for the case of $|m|$ decreasing), a resonance
appears in the energy gap dependence on $R$. The same manifests itself in a sharp change
of the chemical potential derivative. As it is seen, there is no essential difference
in resonance magnitudes of panels (a1) and (b1) of Fig.\ref{fig1}. However, the larger electron
density, the larger resonance density: the distance between two neighboring resonances is
roughly proportional to $1/k_F$ with $k_F$ the Fermi wavenumber of the bulk material.
It is interesting that the resonance effect changes the bulk BCS relation $\Delta_{bulk}
/(k_B T_c) \approx 1.76$ with $T_c$ the superconducting temperature. The ratio
$\Delta_R /(k_B T_{c,R})$ (with $T_{c,R}$ the $R-$dependent transition temperature) is not
constant any more but depends on the width. For example, for the resonances at $R=0.86\;nm$ and
$R=1.12\; nm$ (the panel (a1) of Fig.\ref{fig1}) we have found $\Delta_R /(k_B T_{c,R}) \approx
1.28$. Hence, to get an idea about $T_{c,R}/T_c$ for $0.8\;nm \leq R \leq 1.8 \;nm$ at $n_e=
3.878\;nm^{-3}$, one can simply multiply the data of Fig.\ref{fig1} (a1) by a factor of about
$1.375$. Situation with panel (b1) of Fig.\ref{fig1} is very similar. Here this factor is close
to $1.66$. Shape resonances in the dependence of the quasiparticle spectrum gap on the wire width
are accompanied by strong increase and significant spatial variations of the order parameter
$\Delta(\rho)$. Even for $R > 5\;nm$ the resonance effect is still considerable, and the
superconducting order parameter differs significantly from the bulk value (see two examples of
resonances in Fig.\ref{fig2} (a)). In agreement with calculations for $R > 10\;nm$~\cite{han}
we observe strong oscillations of the order parameter with the frequency roughly proportional to
$k_F$. These oscillations are washed out at large radii about $100\;nm$~(see the results of
Ref.~\cite{han}).

As it is shown, $T_{c,R}$ sharply rises at a resonant point (more than by order of magnitude) and
drops down to a value about $T_c$ between resonances. This makes it possible to expect that the
effect will survive for fluctuating width in a smoothed variant depending on the scenario of
fluctuations. Here it is necessary to mention the increase of the superconducting temperature by a
factor of $1.1$ for a single-crystal $Sn$ nanowire with $R = 10\;nm$~\cite{tian}. Numerical investigation
of Eqs.~(\ref{BdG_simp1}) and (\ref{BdG_simp2}) for the $Sn$ parameters ($g N(0)=0.25$ and $\hbar
\omega_D/k_B= 195\, ^o\!K$, see Ref.~\cite{fetter}) indeed shows the presence of resonances with
$\Delta_R/\Delta_{bulk} \approx 1.1$ near the point $R=10\;nm$. One of them is situated at
$R=9.642\;nm$, where $\Delta_R/\Delta_{bulk} = 1.13$. The superconducting order parameter at this
resonant point has still significant spatial variations in the transverse direction and goes well
above the bulk value (see Fig.\ref{fig2}~(b)). It is worth noting that when exploring the BdG
equations for nanowires with $R > 10\;nm$, the authors of Ref.~\cite{han} have found $10\%-20\%$
deviations of the averaged order parameter from the bulk limit even with radius fluctuations taken
into account.

In summary we investigated influence of the confinement on the transverse superconducting modes in a
cylindrical nanowire at $R < 10\;nm$ in the clean limit. Numerical investigations of the BdG equations
revealed a sequence of significant shape resonances in the dependence of the energy gap parameter on
the nanowire width. The resonant deviations of $\Delta_R$ from the bulk value are by about order
of magnitude larger than the shape resonances in the energy gap parameter of ultrathin metallic films
of the same width. This makes it possible to expect that some smoothed but still profound effect will
remain even in the presence of fluctuations of the wire radius (much less significant resonances in
the superconducting order parameter survive for a nanowire with the average radius $R \approx 10\;nm$
for the variations $\delta R < 0.5\;nm$~\cite{han}).

\begin{acknowledgments}
We would like to thank A. Vagov for useful discussions and remarks.
\end{acknowledgments}


\end{document}